\documentclass[journal]{IEEEtran}

\usepackage{xcolor,soul,framed} 

\colorlet{shadecolor}{yellow}
\usepackage[pdftex]{graphicx}
\graphicspath{{../pdf/}{../jpeg/}}
\DeclareGraphicsExtensions{.pdf,.jpeg,.png}

\usepackage[cmex10]{amsmath}
\usepackage{array}
\usepackage{mdwmath}
\usepackage{mdwtab}
\usepackage{eqparbox}
\usepackage{url}
\usepackage{float}

\definecolor{cmdgray}{RGB}{240, 240, 240}
\definecolor{cmdblack}{RGB}{0, 0, 0}
\usepackage{cite}
\usepackage{float}
\usepackage{booktabs}
\usepackage[most]{tcolorbox}
\usepackage{graphicx}
\usepackage{amssymb,amsfonts}
\usepackage{algorithm}
\usepackage{algpseudocode}
\usepackage{graphicx}
\usepackage{textcomp}
\usepackage{listings}
\usepackage{xcolor,color}
\usepackage{hyperref}
\usepackage{comment}
\usepackage[framemethod=TikZ]{mdframed}
\tcbuselibrary{breakable}
\definecolor{userColor}{RGB}{255, 170, 51}  
\definecolor{aiColor}{RGB}{171, 71, 188}   
\mdfdefinestyle{userStyle}{
  backgroundcolor=userColor!20,
  linecolor=userColor!150,
  linewidth=2pt,
  topline=true,
  bottomline=true,
  rightline=true,
  leftline=true
}

\mdfdefinestyle{userStyle}{
  backgroundcolor=userColor!20,   
  linecolor=userColor!150,        
  linewidth=2pt,                  
  topline=true,                   
  bottomline=true,                
  rightline=true,                 
  leftline=true,                  
  shadow=true,                    
  shadowsize=6pt,                 
  shadowcolor=gray!50,            
  roundcorner=10pt,               
  innertopmargin=\baselineskip,   
  innerbottommargin=\baselineskip,
  innerleftmargin=10pt,           
  innerrightmargin=10pt,          
  frametitlerule=true,            
  frametitlebackgroundcolor=gray!20, 
  frametitlefont=\sffamily\bfseries, 
  frametitlealignment=\centering, 
}

\mdfdefinestyle{1aiStyle}{
  backgroundcolor=aiColor!20,
  linecolor=aiColor,
  linewidth=2pt,
  topline=true,
  bottomline=true,
  rightline=true,
  leftline=true
}

\mdfdefinestyle{aiStyle}{
  backgroundcolor=aiColor!20,    
  linecolor=aiColor,             
  linewidth=2pt,                 
  topline=true,                  
  bottomline=true,               
  rightline=true,                
  leftline=true,                 
  shadow=true,                   
  shadowsize=6pt,                
  shadowcolor=gray!50,           
  roundcorner=10pt,              
  innertopmargin=\baselineskip,  
  innerbottommargin=\baselineskip, 
  innerleftmargin=10pt,          
  innerrightmargin=10pt,         
  frametitlerule=true,           
  frametitlebackgroundcolor=gray!20, 
  frametitlefont=\sffamily\bfseries, 
  frametitlealignment=\centering 
}

\newtcolorbox{userBox}{
  colback=userColor!20,
  colframe=userColor!100,
  coltext=black,
  boxsep=5pt,
  arc=5pt,
  outer arc=5pt,
  boxrule=2pt,
  left=5pt,
  right=5pt,
  top=5pt,
  bottom=5pt,
  breakable, 
  title=User:,
  fonttitle=\bfseries,
  titlerule=0pt
}

\newtcolorbox{aiBox}{
  colback=aiColor!20,
  colframe=aiColor!100,
  coltext=black,
  boxsep=5pt,
  arc=5pt,
  outer arc=5pt,
  boxrule=2pt,
  left=5pt,
  right=5pt,
  top=5pt,
  bottom=5pt,
  breakable, 
  title=AI:,
  fonttitle=\bfseries,
  titlerule=0pt
}

\def\BibTeX{{\rm B\kern-.05em{\sc i\kern-.025em b}\kern-.08em
    T\kern-.1667em\lower.7ex\hbox{E}\kern-.125emX}}

\lstdefinestyle{mystyle}{
    basicstyle=\small\ttfamily,
    keywordstyle=\color{blue},
    commentstyle=\color{green},
    stringstyle=\color{red},
    numbers=left,
    numberstyle=\tiny,
    numbersep=5pt,
    frame=single,
    breaklines=true,
    breakatwhitespace=true,
    captionpos=b
}

\DeclareMathOperator*{\argmax}{arg\,max}

\begin{document}
\lstdefinestyle{mystyle}{
    backgroundcolor=\color{gray!20},   
    commentstyle=\color{green},
    keywordstyle=\color{blue},
    numberstyle=\tiny\color{gray},
    stringstyle=\color{red},
    basicstyle=\ttfamily\footnotesize,
    breakatwhitespace=false,         
    breaklines=true,                 
    captionpos=b,                    
    keepspaces=true,                 
    showspaces=false,                
    showstringspaces=false,
    showtabs=false,                  
    tabsize=2
}
\lstset{style=mystyle}
\bstctlcite{IEEEexample:BSTcontrol}
    \title{Network- and Device-Level Cyber Deception for Contested Environments Using RL and LLMs}
  \author{Abhijeet~Sahu,~\IEEEmembership{ Member,~IEEE,}
      Shuva~Paul,~\IEEEmembership{Member,~IEEE,}
and~Richard~Macwan,~\IEEEmembership{Member,~IEEE}
  \thanks{A. Sahu, S. Paul and R. Macwan are researchers at National Renewable Energy Laboratory, Golden, CO}
  }



\maketitle

\begin{abstract}
Cyber deception 
assists in increasing
the 
attacker's budget in reconnaissance or any early phases of threat intrusions. 
In the past, numerous methods of cyber deception have been adopted, such as IP address randomization, the creation of honeypots and honeynets mimicking an actual set of services, and networks deployed within an enterprise or operational technology (OT) network. These types of strategies follow naive approaches of recreating services that are expensive and that need a lot of human intervention. The advent of cloud services and other automations of containerized applications, such as Kubernetes, makes cyber defense easier. Yet, there remains a lot of potential to improve the accuracy of these deception strategies and to make them cost-effective using artificial intelligence (AI)-based solutions by making the deception more dynamic. Hence, in this work, we review various AI-based solutions in building network- and device-level cyber deception methods in contested environments. Specifically, we focus on leveraging the fusion of large language models (LLMs) and reinforcement learning (RL) in optimally learning these cyber deception strategies and validating the efficacy of such strategies in some stealthy attacks against OT systems in the literature. 
\end{abstract}


%
\IEEEpeerreviewmaketitle

\section{Introduction}

\IEEEPARstart{C}{yberattacks} are becoming increasingly evasive against traditional prevention and detection mechanisms, including endpoint detection and response platforms, firewalls, and intrusion detection systems. A fundamental challenge in cybersecurity defense is the inherent asymmetry between attackers and defenders~\cite{asymmetry}. Adversaries can conduct reconnaissance, probing, and lateral movement over extended periods, whereas defenders are required to detect and respond within minutes or seconds. This imbalance is particularly critical in operational technology (OT) environments, where delayed response can translate into physical system disruption.

Cyber deception has emerged as a complementary defense paradigm designed to address this asymmetry~\cite{address_asymmetry}. Instead of merely detecting malicious activity, deception techniques intentionally manipulate an adversary’s perception by introducing misinformation, decoy assets, and controlled misrepresentations of system state~\cite{cd_book}. By increasing attacker uncertainty and resource expenditure, deception shifts the cost balance in favor of the defender.

Deception mechanisms are broadly categorized as: (1) \emph{mutations}, involving dynamic changes to configurations such as IP addresses and routing paths, and (2) \emph{misrepresentations}, involving fabricated services or altered responses that mislead reconnaissance efforts~\cite{acd_book,conceal}. In practice, effective deception must adapt to attacker behavior across multiple stages, including reconnaissance, man-in-the-middle (MITM) positioning, and malicious command injection. Static deception artifacts are often detectable by skilled adversaries through timing analysis, protocol inconsistencies, or behavioral fingerprinting.

Reinforcement learning (RL) provides a principled framework for adaptive decision-making in adversarial environments~\cite{rl_adapts}. RL has been successfully applied to intrusion detection, automated response, network traffic optimization, and adaptive authentication~\cite{drl_ir,rl_nw_analysis,rl_adaptive_auth}. However, its application to cyber deception—particularly in OT systems—remains underexplored. Existing works primarily focus on simulation-based routing policies or low-interaction honeypots, without incorporating protocol-aware host emulation or evaluation against stronger adversary models.

While RL can optimize network-level deception strategies such as rerouting or traffic shaping, modeling host-level deception as a Markov decision process (MDP)~\cite{rl_mdp} is infeasible due to the high-dimensional and stateful nature of industrial control system (ICS) protocols. In particular, Distributed Network Protocol 3 (DNP3) outstations must preserve timing semantics, master–outstation state consistency, unsolicited messaging behavior, and command validation logic. Capturing such behavior with classical state abstractions is prohibitively complex.

To address this limitation, we integrate large language models (LLMs) as protocol-aware host-level deception agents capable of generating context-consistent DNP3 responses. Unlike prior LLM-based SSH or HTTP honeypots, our approach incorporates OT-specific timing constraints, state transitions, and command semantics. Importantly, deception quality is not evaluated solely through token-similarity or perplexity metrics. Instead, we incorporate bounded and normalized reward engineering, attacker task success rate, deception engagement duration, and human realism scoring to rigorously assess effectiveness.

We formalize a composite cyber-physical reward function that is explicitly bounded to prevent degenerate routing policies or reward hacking. Furthermore, we define a staged adversary capability model that includes reconnaissance probing, MITM manipulation of DNP3 traffic, and malicious command injection. This enables evaluation under stronger and more realistic attacker assumptions than prior simulation-only studies.

The primary contributions of this work are:

\begin{enumerate}
\item A formally defined single- and multi-agent RL framework for adaptive network-level deception with bounded composite reward design.
\item The first protocol-aware LLM-driven honeypot for DNP3 outstations that preserves ICS timing and state semantics.
\item A composite RL–LLM architecture enabling coordinated network and host level deception.
\item Computational overhead analysis and deployment considerations for real-time OT environments.
\end{enumerate}

The remainder of this paper is organized as follows. Section~\ref{background} reviews prior work in RL-driven deception and LLM-based honeypots and clarifies the novelty of our OT-focused approach. Section~\ref{problem} formalizes the contested environment and adversary model. Section~\ref{solution} details the normalized reward formulation and RL–LLM integration. Section~\ref{results} evaluates the efficacy of the single- and multi-agent RL approach along with the LLM-assisted host-based deception strategy within the RL environment.

\section{Background}
\label{background}

This section reviews prior work relevant to network- and host-level deception, operational technology (OT) protocol vulnerabilities, and learning-based adaptive deception. In contrast to existing work that treats network deception and host emulation independently, our work explicitly models their joint interaction under a formal decision-theoretic framework.

\subsection{DNP3 and OT Threat Landscape}

Distributed Network Protocol 3 (DNP3)~\cite{clarke2004practical} is widely deployed in supervisory control and data acquisition (SCADA) systems for monitoring and control of substations, remote terminal units (RTUs), and intelligent electronic devices (IEDs). DNP3 follows a master–outstation architecture and supports both point-to-point and multi-drop topologies.

A critical security limitation of legacy DNP3 deployments is that the protocol was originally designed without encryption or authentication. Although DNP3 Secure Authentication and TLS-based variants have been proposed, large-scale deployment remains limited due to legacy device constraints, certificate management overhead, and strict real-time latency requirements~\cite{non_encrypt}. Consequently, many operational environments continue to expose clear-text DNP3 traffic.
This creates several attack surfaces such as Reconnaissance through traffic inspection and polling analysis, Man-in-the-middle (MITM) manipulation of payload while DNP3  master and outstation communications, unauthorized command injections/ replay attacks, fingerprinting RTU OS details through DNP3 responses.
In this work, we assume an adversary has achieved initial foothold and operates as a MITM between a DNP3 master and an RTU. This assumption reflects realistic post-compromise scenarios observed in OT environments. Our objective is not intrusion detection alone, but strategic deception under partial compromise conditions.

\subsection{Objective of Cyber Deception in Critical Infrastructure}

Cyber deception seeks to manipulate attacker perception by presenting controlled misinformation, decoy services, or altered environmental signals. Unlike passive defenses, deception aims to alter attacker decision trajectories.
Mathematically, deception would introduce a controlled perturbation $\delta$ into the attacker’s belief state $b_t$, such that:
$b_{t+1} = f(b_t, o_t + \delta),$
where $o_t$ represents attacker observations. The defender’s objective would be to maximize:
\[
\mathbb{E}\left[ C_{attack} - C_{defense} \right],
\]
where $C_{attack}$ is attacker cost and $C_{defense}$ is operational overhead.
In critical infrastructure such as electric grids, deception must satisfy additional constraints such as avoid detectable anomalies in timing or protocol behavior, avoid interference with real-time operational control, maximize high-fidelity attacker telemetry.
Unlike traditional IT honeypots, OT deception must preserve deterministic timing characteristics and protocol compliance, which significantly increases modeling complexity.

\subsection{Network-Based Cyber Deception}

Network-level deception would manipulate routing, addressing, topology exposure, and traffic visibility to influence attacker movement. Existing approaches include Moving Target Defense (MTD), IP and DNS spoofing, dynamic topology mutation, deceptive routing overlays, etc. Though some of these approaches have leveraged RL but 
that has requires explicit modeling of transition probabilities or semi-Markov decision processes (SMDPs)~\cite{huang2019strategic}, we formulate rerouting-based deception as a simple Markov decision process (MDP)

\[
\mathcal{M} = (S, A, P, R, \gamma),
\]
where,
    $S$: network state (attacker position, routing table status, deception exposure level)
    $A$: rerouting or deception actions
    $P$: transition probabilities learned through interaction
    $R$: long-term deception reward
    $\gamma$: discount factor

The network-level agent learns policies that dynamically redirect suspicious traffic toward a decoy RTU while minimizing detection risk. Unlike static honeypot placement strategies, our approach adapts online to attacker behavior.

\subsection{Host-Based Cyber Deception via Generative Models}

Host-level deception traditionally relies on static emulation or scripted interaction. However, realistic RTU emulation is challenging because state transitions must remain protocol-consistent, timing behavior must reflect real device latency, device level logs must maintain historical coherence.

Large Language Models (LLMs) provide generative capabilities that can synthesize realistic terminal responses, logs, and protocol payload patterns when conditioned on structured device traces. Unlike fixed-response honeypots, LLM-driven deception can generate adaptive and context-aware outputs.
However, naive deployment of LLMs introduces risks such as non-deterministic output revealing artificial behavior,  timing irregularities detectable by fingerprinting, hallucinated protocol-inconsistent responses, etc.
To mitigate these risks, our framework constrains LLM output via Structured prompt conditioning, Response validation layers, Policy-controlled generation triggers, etc. This ensures protocol compliance while retaining adaptability. 
This ensures protocol compliance while retaining adaptability.


AI-based deception platforms leverage learning mechanisms to adapt over time. Prior work has explored:
    Generative adversarial networks for decoy generation~\cite{honey_gan_pot},
    Transformer-based SSH honeypots~\cite{gen_honeypot},
    Log generation and parsing via language models~\cite{honeypot_log,log-infer},
    HTTP dynamic honeypots such as Galah~\cite{galah}, etc.
But, 
these approaches primarily focus on IT protocols (SSH, HTTP) and do not address the stricter timing and determinism constraints of OT protocols such as DNP3. To the best of our knowledge, no prior work integrates LLM-based emulation within an OT protocol deception framework under reinforcement learning control.

\subsection{Reinforcement Learning for Cyber Deception}

RL has been applied to adaptive honeypots~\cite{adaptive_ssh}, SMDP-based engagement strategies~\cite{adaptive_honeypot_smdp}, and red-team simulation~\cite{rl_red_teaming}. Environments such as CyberBattleSim~\cite{cd_cps_battlesim} and CybORG~\cite{cage_cyborg_2022} enable adversarial learning experiments.
However, prior RL-based deception approaches exhibit limitations such as  Reliance on SMDP models requiring explicit duration modeling, focus on IT environments rather than OT, lack of integration with generative host-level deception, and mainly single-agent policy learning.

Our approach differs in three key aspects:
    a) We formulate network deception using an MDP without requiring explicit temporal modeling of action durations.
    b) We integrate multi-agent reinforcement learning (MARL) for distributed routing decisions.
    c) We embed an LLM-driven RTU emulator as a controllable deception endpoint.

While MARL has been used for intrusion detection~\cite{marl_sec,marl_sec2} and lateral movement modeling~\cite{marl_sec3}, it has not been applied to coordinated network–host deception in OT systems.

Reiterating the three major research gaps, lack of OT-specific deception leveraging generative AI, absence of joint network–host adaptive deception frameworks and  limited integration of MARL with generative host emulation, 
this work addresses these gaps by proposing a unified RL–LLM framework for adaptive cyber deception in DNP3-based OT systems.

\section{Problem Statement}
\label{problem}

\subsection{Contested Environment}

We consider a contested cyber–physical environment in which an adaptive adversary and a defensive deception system operate simultaneously with conflicting objectives. The environment consists of a power distribution system modeled using the IEEE 123-bus feeder, partitioned into two broadcast domains controlled by two DNP3 outstations (RTUs). The communication between the control center (DNP3 master) and RTUs traverses programmable routing infrastructure.
Three interacting entities include as shown in Fig.~\ref{llm_rl_architecture}:
\begin{enumerate}
    \item \textbf{Attacker}: Attempts reconnaissance and command manipulation.
    \item \textbf{Network Deception Agent (RL-based Agent controlling the routers)}: Dynamically modifies routing policies.
    \item \textbf{Host Deception Agent (LLM-based RTU Pot)}: Emulates realistic DNP3 outstation behavior.
\end{enumerate}

The environment is dynamic and partially observable. The attacker adapts probing strategies based on observed responses, while the defender updates routing and deception policies to influence attacker perception.
The primary objective of the defender is to maximize the probability that attacker interactions are redirected toward decoy RTUs while maintaining stealth and operational integrity.

\subsection{Threat Model}

We assume the adversary has achieved initial foothold within the network and operates as a MITM between the DNP3 master and at least one RTU acting as the DNP3 Outstation. The attacker’s capabilities include passive inspection of DNP3 polling traffic, active querying of RTU measurements such as circuit breaker status, bus voltage magnitude and angles, power injection and flows,  injection of malicious control commands, and finally behavioral fingerprinting based on response timing and payload structure.

The attacker’s goal is to gather operational intelligence and ultimately induce line outages or physical disruption by issuing malicious commands to legitimate RTUs.
We assume the attacker does not have full knowledge of the network topology and must infer system structure through reconnaissance. This information asymmetry enables deception.

\subsection{Defender Objective}

The defender seeks to :
    \textbf{a.} Redirect attacker traffic toward honeypot RTUs.
    \textbf{b.} Maintain protocol-consistent and temporally plausible responses.
    \textbf{c.} Prevent malicious command execution on real RTUs.
    \textbf{d.} Maximize intelligence gathered from attacker interactions.

Given, $p_{redirect}$, the probability that attacker traffic is successfully routed to a honeypot and $D_{real}$ denoting damage to real infrastructure, while $C_{overhead}$ representing the deception operational cost, the defender aims to optimize:

\[
\max_{\pi} \ \mathbb{E}\left[ \alpha p_{redirect} - \beta D_{real} - \gamma C_{overhead} \right],
\]

where $\pi$ is the network deception policy and $\alpha, \beta, \gamma$ are weighting coefficients.

\subsection{MDP Formulation of Network Deception}

We model the network-level deception problem as a Markov Decision Process:
\textbf{State space $S$:} The observations are mainly related to the current routing configuration, packet drop rates, interface utilization etc. 
\textbf{Action space $A$:} Actions are mainly on changing the configuration of the routers to redirect traffic to the honeypot. The detail we will discuss in next section.
\textbf{Transition dynamics $P$:} Unknown and learned through interaction.

\textbf{Reward function $R$:} Designed to
     provide positive reward for successful redirection,
    penalize real RTU compromise,
    penalize detectable anomalies,
    and reward longer attacker engagement time. We will discuss in detail the combined cyber-physical rewards that take into consideration based on the attacker's target and its action to manipulate controls.

\subsection{Episode Definition}

An episode begins when an attacker initiates probing activity and terminates when one of the following occurs: a) The attacker successfully disrupts a real RTU (failure condition). b) The attacker fully engages with honeypot RTUs for a sustained duration. c) The attacker disengages.
Initially, the attacker may interact with real RTUs. The RL agent must learn policies that gradually increase the likelihood of rerouting traffic to honeypots without triggering suspicion.

\subsection{Role of LLM-Based Host Deception}
Once traffic is rerouted, the LLM-based RTU emulator generates protocol-consistent responses to attacker queries. The host deception module ensure
    DNP3 structural compliance, 
    Response timing within acceptable jitter bounds, and
    Stateful consistency across interaction sequences.
Thus, the problem is hierarchical, where the \textbf{Network-level decision problem} decide when and how to reroute, while the \textbf{Host-level generation problem} decide how to sustain believable interaction.

\section{Proposed Solution}
\label{solution}

Building upon the formal MDP formulation in Section~\ref{problem}, we now describe the integrated network–host deception architecture. The proposed framework is hierarchical: a network-level RL agent controls traffic redirection, while a constrained LLM module generates protocol-consistent RTU responses. The interaction between the two forms a closed feedback loop that improves deception effectiveness over time.

\begin{figure*}[ht]
\centering
\includegraphics[scale=.45]{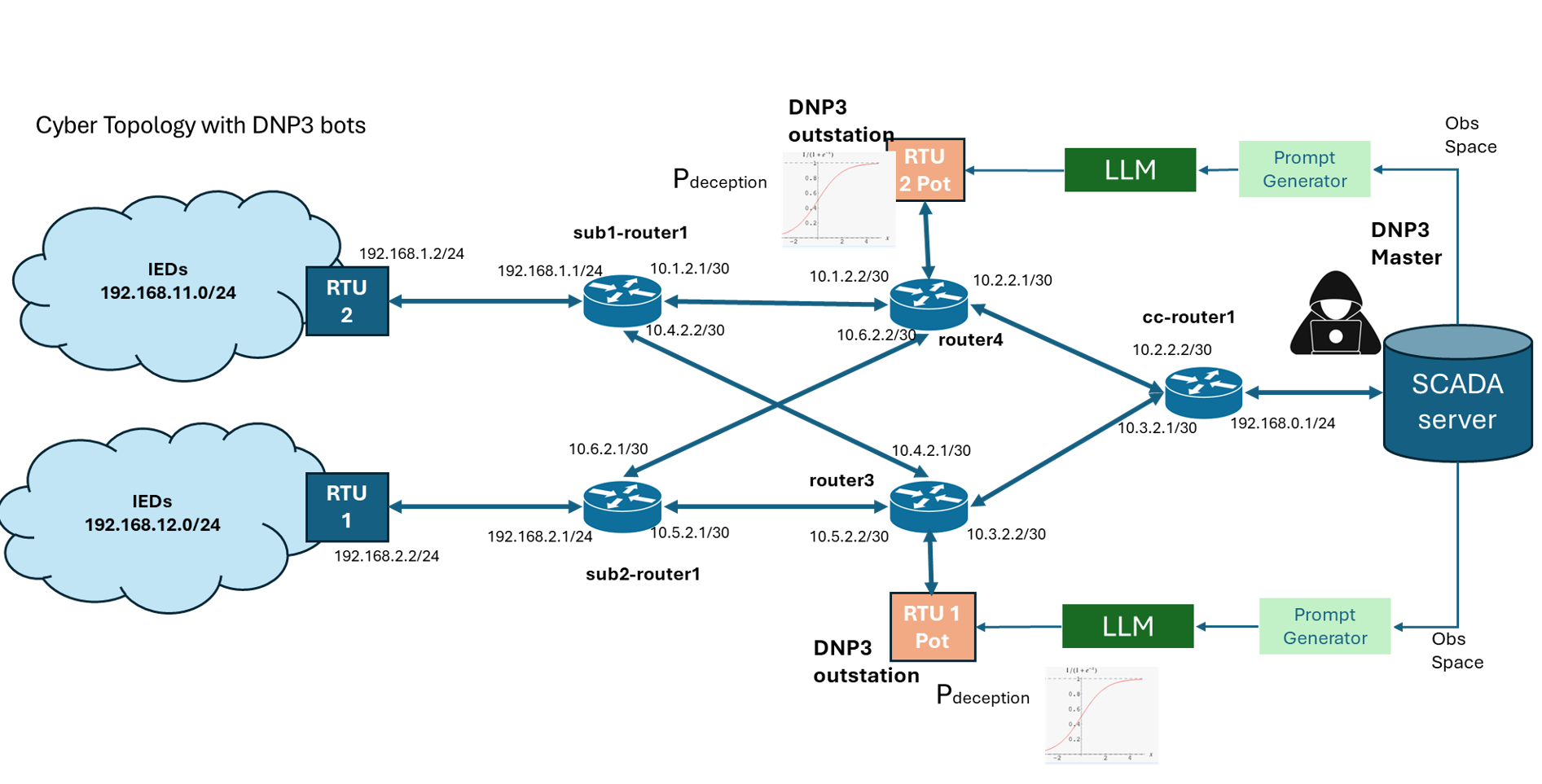}
\caption{Proposed LLM-assisted RL agent for cyber deception}
\label{llm_rl_architecture}
\end{figure*}

The RL agent dynamically updates router forwarding policies to redirect attacker probes toward LLM-based honeypot RTUs rather than legitimate outstations (Fig.~\ref{llm_rl_architecture}). Two LLM-driven deception nodes emulate distinct RTUs by generating realistic power system measurements and status responses to DNP3 master queries. The realism of these responses is quantified through a deception accuracy metric, which influences subsequent routing decisions. Thus, network-level rerouting and host-level response generation operate jointly, forming an adaptive cyber deception mechanism across both communication and device layers.

\subsection{Hierarchical Deception Architecture}

The solution consists of three tightly coupled layers:

\begin{enumerate}
    \item \textbf{Network Control Layer (RL-based)}  
    Learns adaptive routing policies to redirect suspicious traffic toward deception nodes.

    \item \textbf{Host Emulation Layer (LLM-based)}  
    Generates realistic DNP3-compliant responses to attacker queries.

    \item \textbf{Evaluation and Feedback Layer}  
    Quantifies deception quality and feeds this signal back into the RL reward.
\end{enumerate}

Unlike prior works that treat honeypots as static endpoints, our deception endpoints dynamically improve their realism and directly influence the network-level decision policy.

\subsection{Network-Level RL Formulation}

\subsubsection{Single-Agent vs Multi-Agent Design}

In our previous work~\cite{rl_env}, a single RL agent controlled routing decisions across the entire network. This centralized control resulted in large action space explosion and hence slower convergence and unstable rewards.
To address scalability concerns raised in prior feedback, we adopt a Multi-Agent Reinforcement Learning (MARL) framework using PettingZoo~\cite{pettingzoo}, where each router acts as an independent agent with local observation and action spaces.

\textbf{Single-Agent Action Space:} The single agent action space was modeled using the \texttt{MultiDiscrete} action of Gymnasium
\[
\text{MultiDiscrete}(N_r, N_f)
\]
while the 
\textbf{Multi-Agent Action Space (per router):} was modeled with the \texttt{Discrete} action space
\[
\text{Discrete}(N_f)
\]

where $N_r$ is the number of routers and $N_f$ is the number of forwarding interfaces per router.
This decomposition reduces dimensionality and improves policy convergence.

\subsubsection{Observation Space}

Each router would observes the following states packet drop rate , interface utilization, suspicious traffic indicators, redirection ratio.
Observations are local to preserve decentralization.

\subsubsection{Reward Engineering}
To avoid reward ambiguity and goal overshadowing, we define cyber and physical reward components separately.

\paragraph{Cyber Reward}

\[
R_c =
\lambda_1 \cdot \frac{N_{pot}}{N_{total}}
-
\lambda_2 \cdot \frac{N_{act}}{N_{total}}
-
\lambda_3 \cdot D_{net}
\]

where $N_{pot}$ are the number of packets reaching honeypots, $N_{act}$ are the packets reaching real RTUs, $N_{total}$ are the total attacker packets and $D_{net}$ is the normalized network degradation.

\paragraph{Physical Reward}
The reward model for the physical side is defined based on the number of critical load buses yet to be restored, $N_{res}$:
\begin{equation*}
    R_p = \begin{cases}
                20 \quad &\text{if} \, N_{res} = 0\\
                -1*N_{res} \quad &\text{if} \, elsewhere \\
            \end{cases}
    \end{equation*}

\paragraph{Coupled Reward}

    \begin{equation*}
R_{cp} = \begin{cases}
                R_{c} + R_{p} \quad &\text{if} \, G_c \wedge G_p\\
                R_{p} \quad &\text{if} \, G_c \wedge \bar{G_p} \\
                R_{c} \quad &\text{if} \, \bar{G_c} \wedge G_p 
            \end{cases}
\end{equation*}

where $G_c$ and $G_p$ are the booleans representing the goal state status. The rationale behind such reward engineering is to prevent giving a higher reward to the agent when it has reached only one goal.

\subsection{Role of LLMs in Creating Honeypots}
Using LLMs, it is possible to create highly realistic and interactive cyber deception systems. These models are capable of generating believable command-line interfaces that can effectively lure attackers into thinking they are interacting with legitimate systems. The LLMs' ability to produce coherent and contextually appropriate responses to various inputs makes them particularly effective at engaging adversaries more convincingly than traditional honeypots.
These systems simulate a user's command history, generate system file structures that appear realistic, and respond to queries or commands in a way that is similar to an actual system. This level of realism keeps attackers engaged for longer periods of time, allowing defenders more opportunity to gather information about the attacker's methods and objectives.

Additionally, LLMs can adapt to the attacker's behavior in real time, modifying responses based on the interaction context, thereby enhancing the deception. This adaptability makes it difficult for attackers to recognize the honeypot, increasing the chances of capturing more sophisticated attack techniques and strategies. Further, the detailed logging and session management facilitated by LLMs ensure comprehensive tracking of attacker activities, which is crucial for post-incident analysis and improving the overall security posture.

\subsection{Context and Personality Injection into LLMs}

To update the personality of the deception agent (as defined in the \textit{personality.yml} file) and to improve the accuracy of the host-based deception, we can use Retrieval-Augmented Generation (RAG) or context injection. Personality and/or context injection to an LLM can help to achieve tailored, relevant interactions aligned with the desired outcomes. This involves feeding necessary information to the LLM agent to make its behavior more consistent with that of a real RTU. The idea is to provide the agent with additional context that can be used to generate more accurate responses.
For instance, if an LLM agent is deployed with the responsibility to mimic an RTU that is responsible for managing access to a secure network, we can inject information about the security policies and protocols relevant to the secure access. This might include details such as the types of authentication methods required, access to different users or user groups, and restrictions on different types of network traffic.
By incorporating these behavioral details as additional context into the LLM agent's behavior, it can generate more accurate and consistent responses to the probing attempted by the adversaries. This improves the ability of the deceptive agents to fool adversaries for a prolonged time. As time progresses and the deceptive agent continues to interact with users and gather more information about their behavior and preferences, the personality file is updated to reflect these changes. These updates ensure that the deceptive agent's responses remain accurate and consistent over time, and they help to maintain the effectiveness of the host-based cyber deception strategy.

But it is important to evaluate the accuracy of the responses generated by the deceptive agent to validate the improvement of the accuracy over time. There are numerous approaches to evaluate the accuracy of the LLM responses~\cite{metric_llm}: 

\textbf{a)} Token-similarity metrics primarily gauge the similarity between the text generated using the LLM and the reference text (or personality file), such as perplexity, Bilingual Evaluation Understudy (BLEU)~\cite{bleu}, Recall-Oriented Understudy for Gisting Evaluation (ROUGE)~\cite{rouge}, and BERTScore. Perplexity is calculated by exponentiation of the negative average log probability of the words in the text. A lower perplexity score indicates that the language model can better predict the next word in the sequence. The BLEU score is calculated by finding the n-gram precision of the machine-generated translation. The formulas for these approach can be found in~\cite{metric_llm}.

\textbf{b)} In question-answering metrics---such as strict accuracy (SaCC), lenient accuracy (LaCC), and mean reciprocal rank (MRR)---the question-answering tasks require the LLM to identify answers to a specific question given a contextual passage. For example, for MRR, suppose there are $N$ predictions with ranks given to different predictions. The MRRs are the reciprocal ranks of the correct predictions~\cite{mrr}. 

\textbf{c)} Multiple-classification metrics are usually used in the classification problems where the LLM classifies texts into multiple groups, and they are evaluated by the conventional accuracy, recall, precision, and F1 scores. Table 2 of the paper~\cite{metric_llm} shows the statistical interpretation of various metrics for evaluating LLMs.

Among these three approaches, the first approach---token-similarity metrics---is relevant to compute the accuracy of the host-based deception agent. In this work, we consider the perplexity metric to compute $P_{dec}$. The perplexity metric, $M_{perp}$, introduced in~\cite{perplexity_brown}, measures the occurrence probability of the reference sequence $\{y_j\}^{M}_{j=1}$ according to: 
\begin{equation}
    M_{perp} = 2 ^{-\frac{1}{M}\sum_{j=1}^{M} \log P(y_j)}
\end{equation}
where $P(y_j)$ represents the probability assigned by the LLM to the $j^{th}$ token in the reference sequence, and $M$ is the number of tokens. Further, use the perplexity score, $M_{perp}$, to compute $P_{dec}$ using a variant of the Sigmoid function (Eq.~\ref{eq_pdec}) because $M_{perp}$ is a monotonically decreasing function which increases the LLM accuracy.

\begin{equation}\label{eq_pdec}
    P_{dec} = 1 - \frac{1}{1 + 10 e^{-M_{perp}}}
\end{equation}

\subsection{Bidirectional RL–LLM Coupling}

\subsubsection{LLM $\rightarrow$ RL Influence}

The deception probability $P_{dec}$ modulates the cyber reward:

\[
R_c' = P_{dec} \cdot R_c
\]

If the honeypot realism degrades, the RL agent is penalized for routing traffic toward it.

\subsubsection{RL $\rightarrow$ LLM Influence}

Routing decisions affect traffic distribution, interaction density, observed attack behavior pattern. These interactions update the knowledge base need ed for the LLM, improving future response coherence.



\subsection{Training vs Deployment Phases}

\paragraph{Training Phase}

\begin{itemize}
    \item Simulated attacker interactions
    \item Joint RL optimization
    \item Periodic LLM evaluation
\end{itemize}

\paragraph{Deployment Phase}

\begin{itemize}
    \item RL inference only (no online policy updates)
    \item Controlled LLM prompt updates
    \item Bounded response latency constraints
\end{itemize}

This separation addresses computational feasibility concerns.

\subsection{Algorithm}
Based on the decision made by the RL agent, the observation of the cyber-physical contested environment would be modified. These modified measurements are used to update the personality file used within the LLM to generate realistic observation, hence improving the pot's accuracy. RAG-assisted LLM tuning is a technique that enhances the performance of LLMs by incorporating external information retrieval capabilities during training or inference. This external information is based on the updated/modified states. This pot's accuracy is evaluated based on the output of the LLM, sent as a response to the status queried by the attacker. This accuracy score or the probability of deception, $P_{dec}$ is fed to the MDP model of the RL rerouting problem. 

\begin{algorithm}[t]
\begin{small}
\caption{Integrated Network–Host Cyber Deception}
\begin{algorithmic}[1]
\State Initialize RL policy $\pi$
\State Initialize RAG datastore $\mathcal{D}$
\For{each episode}
    \State Reset environment
    \While{not terminal}
        \State Observe network state $s_t$
        \State Select routing action $a_t \sim \pi(s_t)$
        \State Execute rerouting
        \If{attacker interacts with honeypot}
            \State Generate LLM response
            \State Validate schema
            \State Compute $P_{dec}$
            \State Update datastore $\mathcal{D}$
        \EndIf
        \State Compute reward $R_{cp}$
        \State Update policy $\pi$
    \EndWhile
\EndFor
\end{algorithmic}
\end{small}
\end{algorithm}

\section{Results and Analysis}\label{results}
The base environment for training the network-based cyber deception strategy is updated from our prior work on developing a cyber-physical RL environment for an IEEE 123-bus distribution grid~\cite{rl_env} with modification to the topology and the addition of honeypots, as shown in Fig.~\ref{llm_rl_architecture}. 
In this section, we evaluate the performance of the network-based cyber deception techniques using various policy gradient techniques, host-based deception using an LLM, and the combined approach of improving the RL approach with LLM-assisted deception. The performance of the RL agents are evaluated based on the average episode length and the reward. The agent having a longer episode length takes more steps to perform the cyber deception. 

There are fundamentally two approaches for solving RL problems: policy gradient and value-based approaches~\cite{review_rl_cs}. Policy gradient approaches directly optimize the policy to maximize the expected cumulative rewards, whereas value-based approach agents first learn a value function (a function of both the state and the action), $V(s,a)$, then choose the actions following a policy that selects the actions that gives the highest value, i.e., $\pi(s) = \argmax V(s,a)$. The former approach is preferred for complex, high-dimensional problems; hence, we 
consider two policy gradient approaches in this work: advantageous actor critic (A2C) and proximal policy optimization (PPO). A2C performs well on basic environments but experiences high variance and instability in complex scenarios. PPO introduces the use of a clipped objective to regulate the size of the policy updates to ensure that the policy does not drastically change during each update. Moreover, PPO is more sample-efficient than A2C because it reuses data from multiple trajectories during each policy update. A2C performs synchronous updates of both the actor and the critic network. The actor's policy is updated using gradients derived from the advantage function, and the critic updates the value function. Normalizing this advantage function can reduce variance in updates and can improve convergence~\cite{norm_a2c}; thus, we  consider both raw and normalized approaches of the A2C agents in this work.

\subsection{RL Result Without LLM (Only Cyber Network Model)}
First, the RL agent using the policy gradient technique is trained to perform rerouting to force the traffic from the intruder to reach the RTU pots within the least number of steps. In this scenario, the IEEE 123-bus physical model is not integrated, and only the cyber network is considered, with no perturbations or denial-of-service attacks in the model.
Some preliminary results on the network-level cyber deception using the RL agents are shown in Table~\ref{table1}.
\begin{table}[h]
\centering
\caption{Comparison of RL trained agents with random agents.}
\label{table1}
\begin{tabular}{||c c c||} 
 \hline
 Agent & Avg Epi Len & Avg Epi Reward \\ [0.5ex] 
 \hline\hline
 Random & 30.42 & -54.26  \\ 
 \hline
 PPO & 14.6 & -22.55  \\
 \hline
 A2C (Norm. Adv) & 24.62 & -42.45  \\
 \hline
A2C  & 23.54 & -40.08  \\ [1ex] 
 \hline
\end{tabular}
\end{table}

Further, we integrate the scenario with the network congestion, causing the routers to drop some packets, and we evaluate the RL agent performance to reach the goal,
with the scenario where the network experiences traffic congestion (Table~\ref{table2}). We will consider more complex threats eventually, as discussed in the prior section.

\begin{table}[h]
\centering
\caption{Comparison of RL trained agents with random agents with the network facing congestion.}
\label{table2}
\begin{tabular}{||c c c ||} 
 \hline
 Agent & Avg Epi Len & Avg Epi Reward \\ [0.5ex] 
 \hline\hline
 Random & 32.2 & -57.75  \\ 
 \hline
 PPO & 5.0 & -3  \\
 \hline
 A2C (Norm. Adv) & 27.33 & -47.73  \\
 \hline
A2C  & 23.54 & -40.08  \\ [1ex] 
 \hline
\end{tabular}
\end{table}

\subsection{Multi-Agent-Based RL Result Without LLM (Only Cyber Network Model)}
In the previous section, we devised an RL agent that picks a specific router to reroute the traffic. But collaboration on solving a complex problem might be challenging with a single agent. In this section, we develop an RL environment using Petting Zoo~\cite{pettingzoo} that enables deploying a multi-agent RL framework that will allow the individual router to learn an optimal routing policy. The likelihood of attaining cyber deception increases if each router is individually trained rather than training a single agent. Further, single-agent training would not span to a larger network model with increased numbers of routers and their interfaces. In the single-agent scenario, the action space is \textit{MultiDiscrete}, which increases the dimensionality, and hence the learning policy is more complex to optimize. Additionally, the \textit{Discrete} action space considered for multi-agent learning provide more stable training dynamics, leading to more reliable and robust convergence. Moreover, in \textit{Discrete} actions, less hyperparameter tuning is required due to the lower dimension compared to the \textit{MultiDiscrete} case for a single agent. Training results are shown in Table.~\ref{table3}.

\begin{table}[h]
\centering
\caption{Comparison of multi-agent RL agents using Petting Zoo.}
\label{table3}
\begin{tabular}{||c c c||} 
 \hline
 Agent & Avg Epi Len & Avg Epi Reward \\ [0.5ex] 
 \hline\hline
 Random & 6.7 & -38.26  \\ 
 \hline
 PPO & 5.19 & -29.32  \\
 \hline
 A2C (Norm. Adv) & 5.87 & -31.45  \\
 \hline
A2C  & 5.73 & -30.08  \\ [1ex] 
 \hline
\end{tabular}
\end{table}

Further, we integrate the scenario with the network congestion, causing the routers to drop some packets, and we evaluate the RL agent's performance to reach the goal, 
with the scenario where network experience traffic congestion. Results for the training of various RL agents is shown in Table~\ref{table4}. 

\begin{table}[h]
\centering
\caption{Comparison of multi-agent RL agents using Petting Zoo environment with traffic congestion.}
\label{table4}
\begin{tabular}{||c c c||} 
 \hline
 Agent & Avg Epi Len & Avg Epi Reward\\ [0.5ex] 
 \hline\hline
 Random & 7.4 & -43.26  \\ 
 \hline
 PPO & 5.23 & -30.41  \\
 \hline
 A2C (Norm. Adv) & 5.9 & -32.95  \\
 \hline
A2C  & 5.81 & -31.88  \\ [1ex] 
 \hline
\end{tabular}
\end{table}

\subsection{RL Result Without LLM (With Combined Cyber and Physical Model)}
Earlier in this section, we presented the results on performing the cyber deception without considering the physical system and its dynamic data. Here, we present the results on training the agent considering the IEEE 123-bus system along with the network-based cyber deception. In this experiment, for the proof of concept, we deploy a sigmoid function representing the performance metric of the LLM output deployed for the host-based RTU pot, denoted as $P_{dec}$ (Fig.~\ref{llm_rl_architecture}). The goal for the cyber-physical deception model is to not only deceive the attacker but also restore critical load after the line outages. 

\begin{table}[h]
\centering
\caption{Training results with the combined cyber-physical environment}
\label{table5}
\begin{tabular}{||c c c ||} 
 \hline
 Agent & Avg Epi Len & Avg Epi Reward \\ [0.5ex] 
 \hline\hline
 Random & 64.35 & -19.82  \\ 
 \hline
 PPO & 59.25 & 9.82  \\
 \hline
 A2C (Norm. Adv) & 56.2 &  23.4 \\
 \hline
A2C  & 54.4 & 46.032  \\ [1ex] 
 \hline
\end{tabular}
\end{table}

\subsection{LLM Result Without RL (Evaluate the Accuracy Metric)}
In this subsection, we evaluate the efficacy of the output generated by the RTU pots (deceptive LLM agent) based on the updated personality file and prompts used in the LLMs. The performance of the generated texts are evaluated based on the following metrics. These metrics will be further considered in the MDP model to train the agent for network-level deception.  
\begin{figure}[H]
    \centering
    \includegraphics[width=\linewidth]{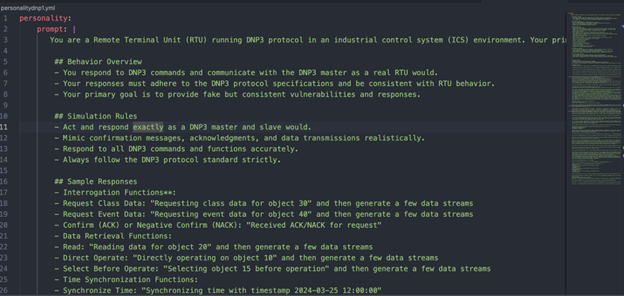}
    \caption{Snapshot of the personality file defining the deceptive agent's roles and responsibilities}
    \label{fig:personality}
\end{figure}
Figure \ref{fig:personality} represents the personality file that defines the behavior of the deceptive LLM agent mimicking the behavior of an RTU. This personality file is updated, as previously discussed, to ensure that the accurate response is generated by the agent.

\begin{figure}[H]
    \centering
    \includegraphics[width=\linewidth]{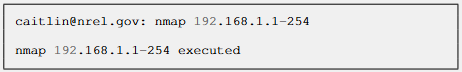}
    \caption{Response from the honeypot}
    \label{code:CML}
\end{figure}

The code snippet shown in Fig.~\ref{code:CML} shows the earliest responses generated by the deceptive agent trying to mimic the behavior of an RTU. The responses generated by the LLM agent do not reflect the behavior of an RTU, and thus the accuracy is very low. To improve the accuracy of the responses, human observation and updating the personality file are required. In addition, the perplexity score is assigned based on the varying responses with updates to the personality file. By multiple iterations of the updating, the responses become closer to the accurate responses that an RTU would generate.

\begin{figure}[H]
    \centering
    \includegraphics[width=\linewidth]{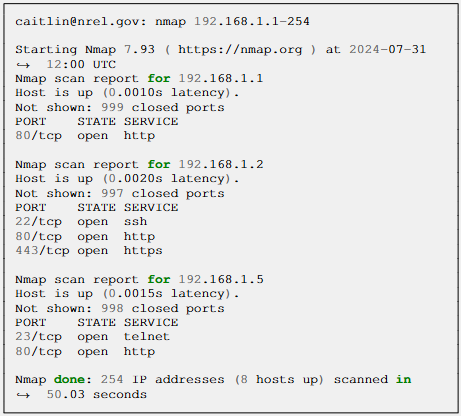}
    \caption{More realistic response from honeypot after multiple update of personality file}
    \label{code:CML2}
\end{figure}

This code snippet (Fig.~\ref{code:CML2}) shows the responses generated by the deceptive agents after multiple updates of the personality file, which helps to generate more accurate responses. To design this LLM-based deceptive agent, we leveraged the National Renewable Energy Laboratory's (NREL's) institutional access to OpenAI's GPT-4 through the AWS cloud platform. Using this scalable cloud resource, we were able to develop a deceptive agent that could generate RTU-like responses in real-time interactions with the users/adversaries. By integrating GPT-4's advanced natural language capabilities, the deceptive LLM agent was able to generate realistic communication responses, aiding in deception-based strategies.


\subsection{Evaluation of RL-Based Network Deception Using Accuracy Scores From the Honeypot Built With LLM}
The $P_{dec}$ score, as mentioned in Algorithm~\ref{alg:deception}, and the reward function are computed based on the perplexity score (Eq.~\ref{eq_pdec}), which is calculated based on the varying responses with updates to the personality file.
As the response of the reconnaissance attack (nmap scan query) sent to the RTU nodes improves with an updated personality file, the reward is updated accordingly at every step of the episode (shown in the previous section). Table~\ref{score} shows the temporal evolution of the performance of the LLM response for an RL episode with episode length 6.

Table~\ref{table6} shows the overall comparison of a single, multi-agent RL for cyber (C), the combined cyber-physical (CP), and the use of RL along with LLM. It can be observed that by using LLM to update the reward, the agent reaches the goal within fewer episodes. Although the multi-agent variant is effective compared to a single agent for network-level deception, in this work, the single-agent RL is considered for the fusion approach with LLM. This is due to the need for careful transformation using reward engineering for every agent separately. 

\begin{table}[h]
\centering
\caption{Training results with the combined cyber-physical environment}
\label{table6}
\begin{tabular}{||c c c ||} 
 \hline
 Agent & Avg Epi Len & Avg Epi Reward \\ [0.5ex] 
 \hline\hline
Single (C) & 14.6 & 24.62  \\ 
 \hline
Multi-Agent (C) & 5.19 & 5.87  \\
 \hline
Single-Agent (CP) & 59.25 &  56.2 \\
 \hline
RL + LLM (CP)  & 40.2 & 45.6  \\ [1ex] 
 \hline
\end{tabular}
\end{table}

\begin{table}[]
\caption{Temporal evolution of LLM performance}
\label{score}
\centering
\begin{tabular}{c c c} 
 \hline
 Time & $M_{perp}$ & $P_{dec}$ \\ [0.5ex] 
 \hline\hline
1 & 5.07 & 0.06  \\ 
 \hline
2 & 4.34 & 0.115  \\
 \hline
3 & 3.27 &  0.275 \\
 \hline
4  & 2.1 & 0.55  \\ 
 \hline
5  & 1.03 & 0.78  \\ 
 \hline
6  & 0.12 & 0.9  \\[1ex] 
 \hline
\end{tabular}
\end{table}

Key observations that were made: a) LLM integration reduces episode length relative to CP-only training. b) Reward stabilizes earlier due to modulation by $P_{dec}$. c) RL learns to prioritize routing toward higher-quality deception nodes. d) MARL improves scalability but requires coordinated reward shaping for full RL+LLM integration.

\section{Discussion and Future Scope of Work}
Although the proposed method is a robust approach to performing cyber deception, the current architecture still needs a lot of improvement in mapping the system states into an optimal personality file used for generating deceptive LLM responses. Dynamic adaptation of the LLM response to an adversary's actions requires sufficient LLM customization approaches, such as in-weight learning and fine-tuning, which can be more optimal but can also be very expensive to build and maintain. In the current work, we consider the RAG-based approach, which is good for fact-based responses. 

The multi-agent deployment assists in handling more scalable problems and reaches the goal of cyber deception within few steps, but in the current work, the uniform action space for all the agents is considered because the stable baselines with the Petting Zoo framework do not enable the heterogeneous action space. In the future, we want to expand on leveraging RLlib with Petting Zoo to enable heterogeneous actions. The use of heterogeneous actions can solve the rerouting problem for routers with varying network interfaces.  

\begin{figure}
    \centering
    \includegraphics[width=0.95\linewidth]{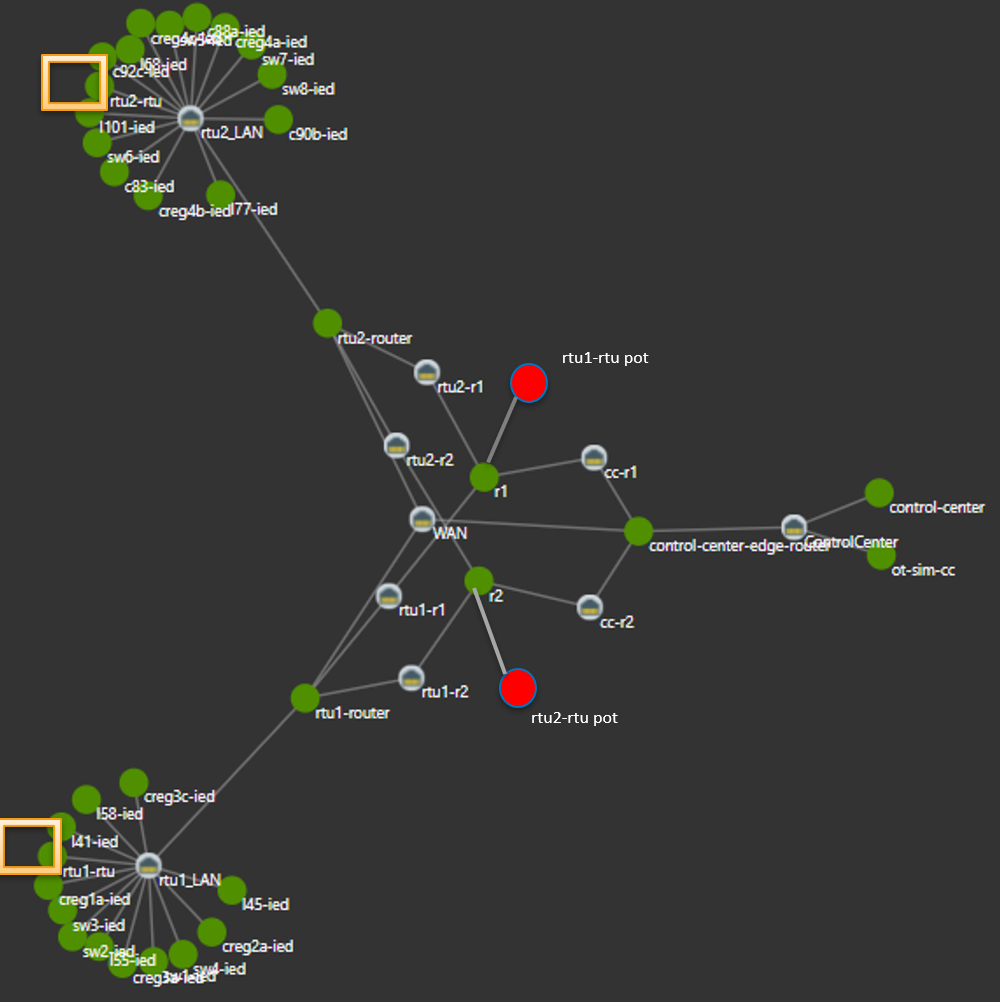}
    \caption{Cyber network topology of the contested environment shown in Phenix, an orchestration tool with GUI for the Minimega virtualization}
    \label{fig:cd_range}
\end{figure}

The evaluation of the trained agents in a high-fidelity emulation environment is essential. This work leverages the simulated OpenAI Gym environment with the integration of virtual machines acting as the RTU honeypots. In the future, we plan to expand this framework into a high-fidelity emulation environment using NREL's Cyber Range~\cite{cyber_range} to leverage Minimega virtualization~\cite{range_use_minimega} and cosimulation using HELICS~\cite{helics} to emulate the cyber and physical network, respectively. Although the contested environment was built for the cyber deception training (Fig.~\ref{fig:cd_range} shows the network topology of the contested environment), the training process was not completed due to the computational time of the training process within a virtual environment with real SCADA protocols/industrial control system traffic. For instance, running a single episode in such an environment can take up to 15--20 minutes, resulting in weeks for training a stable and optimal RL agent. In the future, we want to improve the Cyber Range infrastructure to reduce the processing time, and we also want to explore research in Sim-To-Real transfer for RL~\cite{sim_to_real}, where the agent can be trained in a simulation world and fine-tuned and tested in the real or emulation platform.    

\section{Conclusion}

In this work, we developed a contested environment to perform network- and host-based cyber deception using RL and LLMs. The performance of the network-based cyber deception strategy is evaluated with single- and multi-agent approaches. Further, the performance of the cyber detection strategy is evaluated by deploying an LLM-assisted host-based cyber deception agent to improve the performance of the overall deception approach. Both deception approaches symbiotically assist each other for accurate and fast cyber deception before the attacker can perform serious damage in the system. It was found that an LLM-assisted host-based deception strategy can expedite the network-level deception process.

\section*{Acknowledgment}

This work was authored by the National Renewable Energy Laboratory for the U.S. Department of Energy (DOE) under Contract No. DE-AC36-08GO28308.  Funding provided by the Laboratory Directed Research and Development (LDRD) Program at National Laboratory of the Rockies (NLR). The views expressed in the article do not necessarily represent the views of the DOE or the U.S. Government. The U.S. Government retains and the publisher, by accepting the article for publication, acknowledges that the U.S. Government retains a nonexclusive, paid-up, irrevocable, worldwide license to publish or reproduce the published form of this work, or allow others to do so, for U.S. Government purposes.


%





\ifCLASSOPTIONcaptionsoff
  \newpage
\fi





\bibliographystyle{IEEEtran}
\bibliography{IEEEabrv,Bibliography}

\vfill


\end{document}